 \documentclass[preprint,12pt]{elsarticle}

\usepackage{amssymb}
 \usepackage{amsmath}
 \usepackage{amsthm}
 \usepackage{amsfonts}

\journal{Chemical Physics}

\begin{document}

\begin{frontmatter}



\title{Spin-Orbit Based Coherent Spin Ratchets}


\author{Matthias Scheid$^1$}
\author{Dario Bercioux$^{2}$}
\author{Klaus Richter$^1$}

\address{$^1$Institut f\"ur Theoretische Physik -- Universit\"at Regensburg, 93040 Regensburg, Germany}
 \address{$^2$Freiburg Institute for Advanced Studies
 and Physikalisches Institut, Albert-Ludwigs-Universit\"at, 79104 Freiburg, Germany}

\begin{abstract}
The concept of ratchets, driven asymmetric periodic structures
giving rise to directed particle flow, has recently been generalized to a quantum
ratchet mechanism for spin currents mediated through spin-orbit interaction.
Here we consider such systems in the coherent mesoscopic regime and
generalize the proposal of a minimal spin ratchet model based on a non-interacting clean
quantum wire with two transverse channels by including disorder and by
self-consistently treating the charge redistribution in the nonlinear (adiabatic) ac-driving regime.
Our Keldysh-Green function based quantum transport simulations show that
the spin ratchet mechanism is robust and prevails for disordered, though
non-diffusive, mesoscopic structures. Extending the
two-channel to the multi-channel case does not increase the net ratchet
spin current efficiency but, remarkably, yields 
a dc spin transmission increasing linearly with channel number.
\end{abstract}


\end{frontmatter}


\section{Introduction}

The appealing physical concept of converting energy from randomly moving Brownian 
particles in asymmetric set-ups into a directed particle flow, possibly against
an external load, has led to an enormous amount of works establishing an
own field at the interface of transport and nonequilibrium statistical physics. 
Peter H\"anggi, as one of the founders of this branch of statistical physics, coined 
the term "Brownian motors" for such systems \cite{Bartussek95}.
Ratchets, spatially periodic structures with broken left-right symmetry
operating far from equilibrium and thereby generating directed particle
motion in the presence of unbiased time-periodic driving
constitute one important class of such systems. The ratchet mechanism
was first discovered for classical Brownian particles
\cite{Haenggi96,Astumian,Juelicher97}
and then generalized to quantum dissipative systems~\cite{Reimann97}.
Later this concept has  been extended to the coherent regime where
corresponding ratchets and rectifiers have gained increasing attention,
in particular after the experimental demonstration of ratchet-induced
charge flow in periodically arranged lateral quantum dots
based on a two-dimensional semiconductor heterostructure \cite{Linke99}.
Coherent rectifiers are characterized by phase-coherent
quantum dynamics in the central periodic
system in between leads where dissipation takes place. For a recent
comprehensive review of the whole, broad field of ratchets and 
artificial Brownian motors
 see \cite{Hanggi09}.

While nearly all the works in this field have addressed the problem
to achieve unbiased directed {\em particle} transport, we have generalized
this concept to the notion of {\em spin} ratchets, corresponding
set-ups which allow for generating spin currents, partly even
in the absence of charge currents. In this respect, "Zeeman ratchets"
that are based on an asymmetric, spatially periodic magnetic field are
closest to the usual charge ratchets: Owing to the Zeeman term in the
Hamiltonian, spin-up and -down electrons experience opposite asymmetric
periodic potentials, which should give rise to a net flow
of charge carriers with different spin polarization in opposite
directions, corresponding to a pure spin current \cite{Scheid2006}.
Similar results have been found for a one-dimensional quantum wire with
strong repulsive electron interactions \cite{Braunecker2007}.
However, contrary to particle ratchets with preserved particle number,
spin-polarization is a volatile property and subject to spin relaxation. Hence in
\cite{Scheid2007a} we showed, both conceptually and numerically for a
realistic model of a Zeeman ratchet including spin-flip processes that
pure spin current generation in coherent mesoscopic conductors is indeed  possible.
In parallel we devised the concept of a "spin-orbit ratchet" \cite{Scheid2007},
a setting where the spin orbit interaction (SOI) in a quantum wire is
employed to generate a spin current.
Contrary to particle ratchets, which rely on asymmetries in either the spatially
periodic modulation or the time-periodic driving, a spin-orbit (SO) based ratchet
works even for symmetric electrostatic periodic potentials, due to the spin-inversion asymmetry
of the SOI. As a result, no ratchet charge current is produced in parallel, leading
to a pure ratchet spin current \cite{Scheid2007}. As possible realizations we have in mind
systems based on 
semiconductor heterostructures with Rashba SOI~\cite{rashba:1960} 
that can be tuned in strength by an external gate voltage allowing to control the spin evolution.

Such a SO-based ratchet is an example of a device for "mesmerizing"
semiconductors \cite{Bauer2004}, i.e.\ for magnetizing semiconductors without
using magnets, in contrast to the usual method of spin injection through ferromagnets.
Also in this sense spin ratchets  have much in common with
 spin pumping \emph{i.e.}\ the generation of spin-polarized currents at zero bias via
cyclic parameter variation. 
Different theoretical proposals based on SO~\cite{Sharma03} and Zeeman~\cite{Mucciolo02}
mediated adiabatic spin pumping in non-magnetic semiconductors have been suggested
and, in the latter case, experimentally observed in mesoscopic cavities~\cite{Watson03}.
Adiabatically driven spin ratchets, however, also differ from adiabatic spin pumps,
 since ratchets operate with a single ac driving parameter in the nonlinear bias regime, 
while the usual proposals for adiabatic pumping include a cyclic variation of at least two
parameters at zero external bias \cite{comment-pumping}.

Recently, the concept of SO-based spin ratchets has been generalized to the
quantum dissipative regime. To this end the above setting was extended
by coupling the orbital degrees of freedom additionally to an external
bath (within a Caldeira-Leggett model \cite{Caldeira-Leggett}) representing,
e.g. effects of phonons. While SO-mediated spin-phonon
coupling usually leads to spin relaxation, it could be shown that for this ratchet
set-up the opposite is true: a finite, pure spin current is generated
\cite{Smirnov08a,Flatte08}. This is remarkable as it means that thermal
energy from the bath is converted, via the SO coupling, into a directed
spin current with aligned charge carrier spins; in this sense such a
dissipative spin ratchet can be viewed as a "Brownian spin motor". An
extension to an additional in-plane magnetic field which allows for
further controlling the spin current can be found in \cite{Smirnov08b}.
The study of dissipative SO ratchets took another twist when it turned
out that a finite {\em charge} current arises in a SO ratchet even
if both the spatially periodic potential and the time-periodic driving
are symmetric \cite{Smirnov09}. While directed transport is well known 
to appear for symmetric potentials as long as the driving contains a time asymmetry
\cite{Goychuk98,Flach00,Yevtushenko00}, it seems paradoxical
that one can even go without this symmetry breaking. The charge ratchet
mechanism for this space- and time-symmetric case results from the
interplay between quantum dissipation and SO-induced spin-flip processes
and thereby from a hidden symmetry breaking through the SO coupling.
This new class of space- and time-symmetric charge ratchets is hence built 
on the intrinsic spin nature of the particles to be transported, exhibiting
certain conceptual similarities with the classical "intrinsic ratchets" discussed 
in \cite{Broek09}.

In the present paper we however focus on coherent SO ratchets and
extend the minimum model introduced in \cite{Scheid2007} in several directions
in order to gain an improved  and systematic understanding of the spin 
ratchet mechanism and its limits
for set-ups describing realistic mesoscopic devices. 

First, most investigations of ratchets disregard interaction
effects and usually model the time-periodic driving in terms of the bare
homogeneous external field. However, in particular for charge ratchets in
the nonlinear bias regime, the actual voltage drop across the system
may deviate from simple bias models often used, e.g.\ from a linear model,
 and hence can affect the
ratchet current which is known to depend sensitively on the details of the system.
Here we will consider in a self-consistent treatment the charge redistribution 
in the ratchet due to the external bias giving rise to a nonlinear voltage drop 
and thereby presumably altering the resulting spin-dependent transmission through the device.

Our study includes, second, an analysis of disorder effects usually unavoidable in
mesoscopic devices.
It is well known that impurity scattering in a SO medium usually leads to Dyakonov-Perel-type 
spin relaxation \cite{Dyakonov1972}. This mechanism will obviously counteract the generation of 
spin-polarized currents, and hence it is important to understand to which extend 
the spin ratchet mechanism is robust against disorder effects.

Third, as shown in \cite{Scheid2007}, at least two SO-coupled
transverse modes in a quasi-one-dimensional wire with at least one electrostatic
potential barrier are required for generating a net spin current. While an
increase in the barrier number on the whole leads to an enhanced ratchet
spin current \cite{Scheid2007}, the spin current dependence on the number $n$
of transverse modes, respectively the wire width, remains to be investigated. 
Our analysis shows that, on average, the ratchet spin current does only marginally 
increase with $n$. However, interestingly, we find for a given (dc) bias a linear 
increase of the spin-polarized transmission with $n$. 

The paper is organized as follows: After introducing the system and the
numerical method in Sec.~2 below, in Sec.~3 we will first consider spin-polarization 
effects in dc-transport through two-dimensional ribbons with Rashba SOI
and outline an important underlying polarization mechanism. In Sec.~4 we
then study the SO ratchet response for the nonequilibrium and disordered case 
and conclude with a number of remarks in Sec.~5.


\section{Outline of the system}
\begin{figure}[!b]
	\begin{center}
 \includegraphics[width=0.8\linewidth]{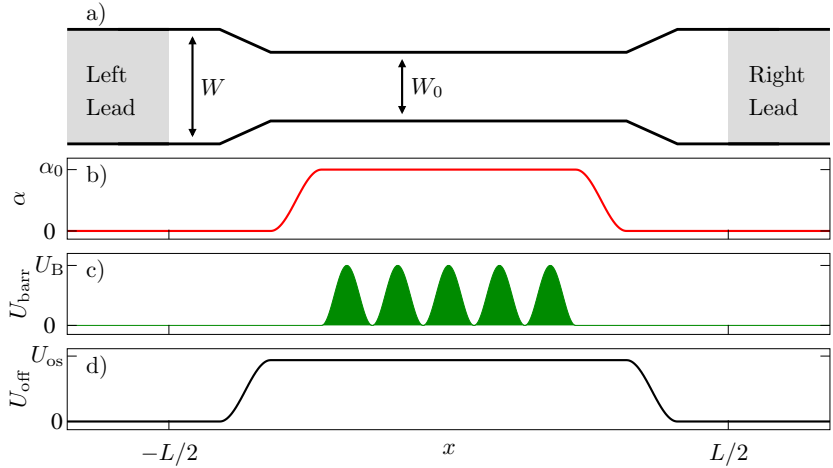}
	\caption{\label{SORat}a) Spin ratchet setup consisting of a quantum wire along the $x$-direction.
The width of the leads is $W$ while the central region has a width $W_0$. b) strength of the Rashba spin-orbit interaction
 $\alpha (x)$; c) electrostatic potential barriers $U_\mathrm{barr}(x)$; d) additional electrostatic potential offset $U_\mathrm{off}(x)$.
	}
	\end{center}
\end{figure}

We consider a quantum wire (oriented in $\hat{x}$-direction) which can be regarded
as being realized in a two-dimensional electron gas (2DEG) in the $(x,y)$-plane. 
A typical structure is visualized in Fig.~\ref{SORat}a.
In the central region of the wire Rashba SOI is present, which  is described by 
the Hamiltonian
%
\begin{equation}\label{HamRashSym}
H_\mathrm{R}=
\frac{1}{\hbar} \alpha(x) \sigma_x p_y -
\frac{1}{2\hbar} \sigma_y \big[\alpha (x)p_x +p_x\alpha (x)\big]  .
\end{equation}
%
To avoid unwanted reflections at the interfaces between the SO-free leads and the scattering region with
finite SOI $\alpha(x)$, we adiabatically turn on the Rashba SOI (see
Fig.~\ref{SORat}b). Since Rashba SOI is only present in the central region of the wire
\cite{Nikolic},
we do not face difficulties with the definition of a spin current inside leads with SOI~\cite{Shi2006}.

In addition we consider $N_\mathrm{B}$ identical potential barriers, which are located in the wire 
(see Fig.~\ref{SORat}c). They are modeled by the electrostatic potential
\begin{equation}\label{Barrier}
U_\mathrm{barr}(x)=
\begin{cases}
\frac{1}{2}
 U_\mathrm{B}\left\{ 1-\cos 
\left[
\frac{2\pi}{ L_\mathrm{B}}\left(x+\frac{1}{2} N_\mathrm{B}L_\mathrm{B}\right)
\right]\right\}&\mathrm{for}\;|x|<N_\mathrm{B}L_\mathrm{B}/2\\0 &\mathrm{elsewhere}
 \end{cases}
\,.
\end{equation}
The Hamiltonian of the whole system then reads
%
\begin{equation}\label{Hsim}
H=\frac{p_x^2 +p _y^2}
{2m^*}+ U_\mathrm{conf}(x,y)+U_\mathrm{barr}(x)+\delta U_\mathrm{es}(x,y)+H_\mathrm{R}\,,
\end{equation}
%
%
where $U_\mathrm{conf}(x,y)$ is the hard-wall confinement potential and $\delta U_\mathrm{es}(x,y)$ is an
electrostatic potential due to the rearrangement of charges at finite bias, see Sec.~4.1.

A quantitative study of nonequilibrium and disorder effects on ratchet transport requires a
numerical approach. For the respective transport calculations, we use a tight-binding 
version of Eq.~(\ref{Hsim}), i.e.\ the system is discretized on a square grid with lattice spacing $a$. 
By using an efficient recursive algorithm for computing the lattice Green functions \cite{WimmerRichter09}
we determine the relevant spin-dependent transport properties of the system. From the numerically
obtained $S$-matrix elements we then calculate the spin resolved quantum transmission 
probabilities $T^{RL}_{\sigma ,\sigma '}$ between the left and right lead. 
Here $T^{RL}_{\sigma ,\sigma '}$ describes the probability for an electron with spin state 
$\sigma '$ to be transmitted from the left entrance lead into the right  exit lead with spin state $\sigma$. 
The spin state $\sigma = \pm$ is defined with respect to a spin quantization axis pointing in 
$y$-direction. Such an in-plane polarization is often referred to as Edelstein effect \cite{Edelstein}.
From the spin-resolved transmissions we can then calculate the total and
spin transmission, respectively:
\begin{eqnarray}\label{Ttotal}
T(E) & = & \sum _\sigma \sum _{\sigma '} T^{RL}_{\sigma ,\sigma '}(E) \, ,  \\
T_\mathrm{S}(E) & = & \sum _{\sigma '} \left[ T^{RL}_{+ ,\sigma '}(E)-T^{RL}_{- ,\sigma
'}(E)\right] \, . 
\label{TS}
\end{eqnarray}
Contrary to the total transmission, Eq.~(\ref{Ttotal}), which in a two-terminal setup is symmetric 
with respect to interchanging the leads, i.e.\ $T=T^\mathrm{RL}=T^\mathrm{LR}$, 
the spin transmission (and naturally also the associated spin current) can differ from lead to lead:
in general $T_{\mathrm{S}}^\mathrm{RL}
\neq T_{\mathrm{S}}^\mathrm{LR}$ owing to the spin inversion asymmetry
of the SOI. Therefore it is necessary to specify the lead where the spin current is evaluated. 
We will calculate the spin transmission from the left to the right lead using the abbreviations
$T_{\mathrm{S}}(E)=T_{\mathrm{S}}^\mathrm{RL}(E)$, see Eq.~(\ref{TS}), 
and 
 $T_{\mathrm{\sigma \sigma'}}(E)=T_{\mathrm{\sigma\sigma'}}^\mathrm{RL}(E)$ employed below.
For later use we furthermore define dimensionless energies and Rashba SOI strengths 
(denoted by a bar):
$\bar{E}=[\hbar ^2/(2m^*a^2)]E$ and $\bar\alpha = (m^* a/\hbar ^2 )\alpha$.
\noindent
\section{Spin polarized dc-transport}
Before turning our attention to the ratchet behavior upon ac-driving we investigate the 
dc-transport properties of the system outlined in Fig.~\ref{SORat} in order
to demonstrate the effect of the SOI underlying the spin ratchet mechanism.
To this end we evaluate the charge/spin currents in the leads in response to a fixed finite, 
applied bias making use of the expressions from the Landauer-B\"uttiker formalism. The 
charge current for coherent transport in a quantum wire reads
%
%
\begin{equation}\label{Ic}
I= \frac{e}{h}\int_{0}^{\infty} \mathrm{d}E \;\big[ f(E;\mu _\mathrm{R}) - f(E;\mu _\mathrm{L} ) \big] \;T(E) \,.
\end{equation}
%
%
Here $f(E;\mu )$ is the Fermi-Dirac distribution function and $\mu _\mathrm{L/R}$ is the chemical potential of the left/right lead. Correspondingly, the spin current in the right lead reads~\cite{Scheid2007a}:
\begin{equation}\label{Is}
I^{\mathrm{S}}=
\frac{1}{4\pi}\int_{0}^{\infty} \!\!\mathrm{d}E\;
\big[ f(E;\mu _\mathrm{L} ) - f(E;\mu _\mathrm{R})  \big] T_{\mathrm{S}}(E)  \, .
\end{equation}

%
\begin{figure}[!tb]
\begin{center}
 \includegraphics[width=0.8\linewidth]{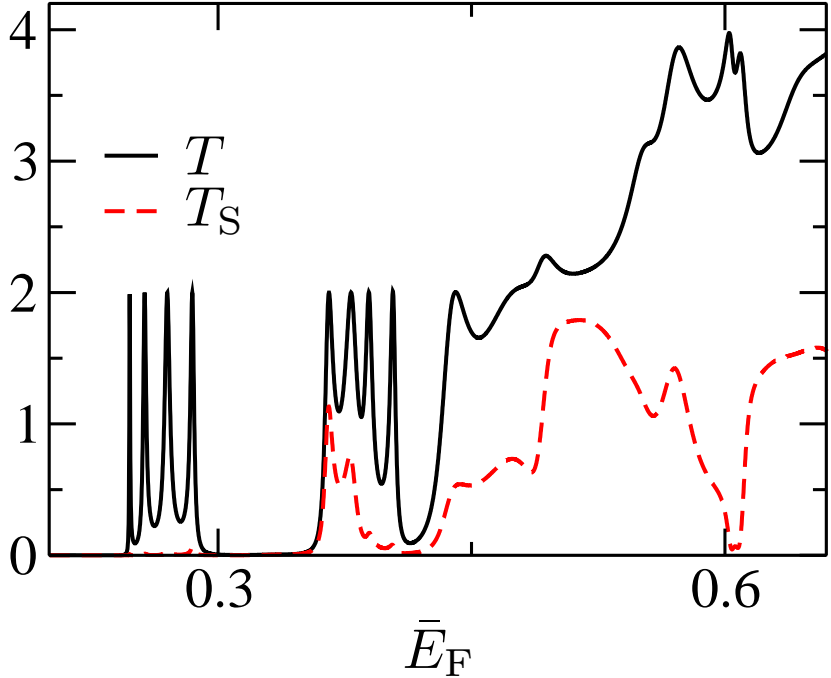}
  \caption{Charge (full black line) and spin (dashed red line) transmission probability 
in linear response for the system depicted in Fig.~\ref{SORat}. 
Parameters used are: $L_\mathrm{B}=10a$, $N_\mathrm{B}=5$, $\bar{\alpha} _0 =0.15$, $W=25a$, $W_0=15a$, $\bar{U}_\mathrm{B}=0.2$, $\bar{U}_\mathrm{os}=0.15$.
  \label{Trans}}
\end{center}
\end{figure}\noindent
In Fig.~\ref{Trans} we show the numerically computed transmission probabilities 
$T(E)$ and $T_\mathrm{S}(E)$ in linear response for a system with five potential barriers,
Fig.~\ref{SORat}.
As shown $T(E)$ (full black line) exhibits two combs of four transmission peaks each 
due to resonant tunneling through the array of potential barriers, reflecting precursors of
minibands arising for an infinite array of barriers. The two transmission combs belong to transport
of states with transverse mode number $n\!=\!1$ and $2$, respectively.  Furthermore, the system exhibits 
strong spin polarization in $+y$ direction, i.e.~$T_\mathrm{S}\ge 0$ (dashed red line), 
for a wide range of Fermi energies
above the subband energy of the second channel. This indicates that SO mixing of at least two
modes is required for a nonzero spin transmission.

\begin{figure}[!bt]
\begin{center}
  	\includegraphics[width=0.8\linewidth]{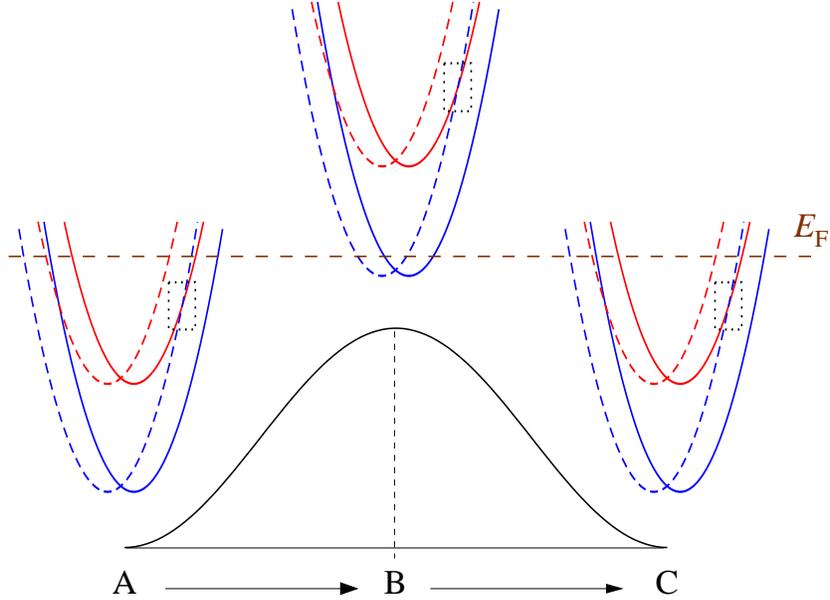}
	\caption{\label{LZMod}Sketch of the parabolic energy dispersion for two transverse modes of the quantum wire in the presence of an electrostatic barrier shown at three different positions (A)-(C). The spin-splitting due to SOI is indicated by the shift of the parabolas in $\pm k_x$-directions. 
The dotted square indicates an avoided crossing of transversal modes $(n,\sigma) = (2,+)$ and
$(1,-)$ with different spin polarization, where the spin-flips can happen, 
see Eq.~(\ref{LandauZener}). The Fermi energy is marked by the horizontal dashed line.
	}
\end{center}\vspace*{-1em}
\end{figure}

We now outline the basic mechanism which causes such a spin polarization. To this end we employ a 
Landau-Zener model for a single barrier which was first introduced by Eto et al.~\cite{Eto2005} 
to describe spin-dependent transport across a quantum point contact (see also \cite{DA-Pfund,Balseiro2007}).
In Fig.~\ref{LZMod} we show the SO-split parabolic dispersions (in $k_x$-direction) of
charge carriers in the two first transversal subbands of the quantum wire,
relative to the fixed Fermi energy $E_\mathrm{F}$ (horizontal dashed line),
at three different positions A-C of the potential barrier (along the $x$-direction).
The SOI further couples different parabola branches and leads to a small avoided crossing 
(position marked by a dashed box) between
the states $(n,\sigma)=(2,+)$ and $(1,-)$ (full red and dashed blue line) as shown 
in Fig.~\ref{LZMod}.
In \cite{Eto2005} it was found that upon traversing the barrier (A$\rightarrow$B$\rightarrow$C), 
first higher transversal modes of the wire become depleted (see position B in Fig.~\ref{LZMod}). 
After passing the barrier top the SOI gives rise to a spin-dependent repopulation of these higher modes 
between position B and C when the Fermi energy passes adiabatically the afore mentioned anticrossing.

\begin{figure}[!tb]
\begin{center}
  \includegraphics[width=0.8\linewidth]{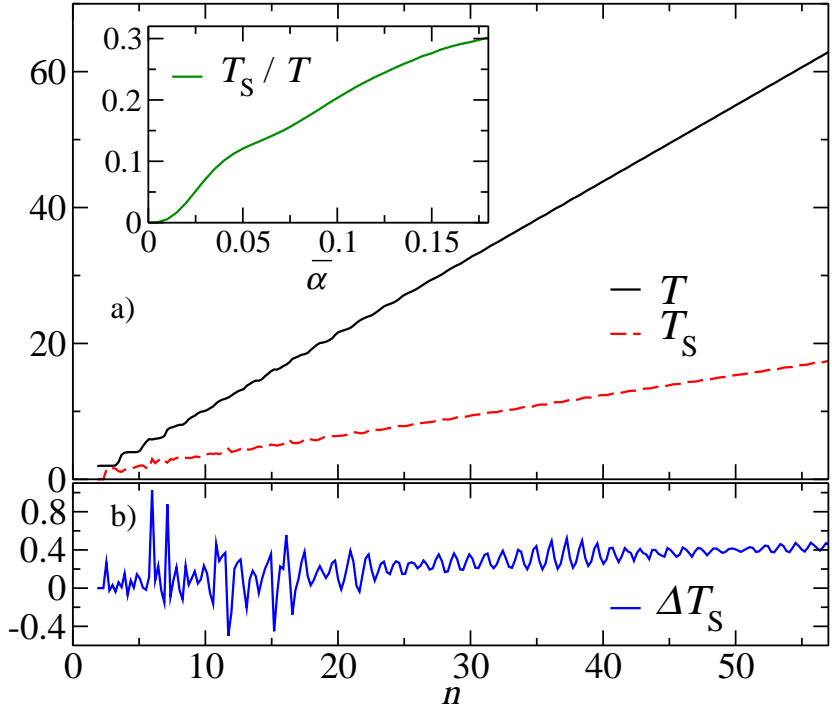}
  \caption{a) Total transmission (full black line) and spin transmission
(dashed red) in linear response as a function of the channel number $n$ in the lead for a scaled 
 SOI strength $\bar{\alpha}_0=0.15$. b) Corresponding ratchet spin
transmission $\Delta T_S$, Eq.~(\ref{ratspintrans}),
for a linear voltage drop at finite bias $\bar{U}_0=0.02$. Inset of panel a): Spin polarization
ratio $T_S / T$ as a function of the scaled SOI strength at fixed width $W=250a$ corresponding to
$n=56$ transverse channels. 
Common parameters: $L_\mathrm{B}=60a$, $N_\mathrm{B}=1$, $\bar{E}_\mathrm{F}=0.36$, $\bar{U}_\mathrm{B}=0.22$, $U_\mathrm{os}=0$.
  \label{HangMan}}
\end{center}
\end{figure}

For the simplest situation of two occupied transversal modes and a single barrier 
(shown in Fig.~\ref{LZMod}) the spin transmission can be estimated as~\cite{DA-Pfund}
\begin{equation}\label{LandauZener}
T_\mathrm{S}=2P\,,\quad \mathrm{where}\qquad 
P=1-e^{-\lambda}
\end{equation}
\noindent
is the transition probability between subbands with different spin polarization 
\mbox{$(n=1,\sigma=-)\leftrightarrow (n=2,\sigma=+ )$}. 
This quantity was evaluated in Ref.~\cite{Eto2005} using Landau-Zener theory~\cite{Landau1932,Zener1932}, 
where $\lambda\ge 0$ parametrizes the adiabaticity of the transition. In the diabatic limit,
$\lambda \rightarrow 0$, Landau-Zener transitions preserving the state $(1,-)$ dominate, while
in the adiabatic limit the $(1,-)$ state (dashed blue line in Fig.~\ref{LZMod}) 
changes its character into $(2,+)$ (solid red line). The Landau-Zener parameter
$\lambda$ depends on the form of the barrier, the confinement potential and the SOI. 
As two of us showed in Ref.~\cite{Wimmer2009}, $T_\mathrm{S}$ increases with the length 
of a
 point contact constriction, which corresponds to the adiabtic limit of an increasing length
$L_\mathrm{B}$ of the barrier in the set-up investigated here.\\
Generalization of these results to a higher number $n$ of contributing transversal channels 
shows that this spin polarization effect is not limited to the two-channel case. 
In Fig.~\ref{HangMan} we present our numerical results for $T$ and $T_\mathrm{S}$ for
fixed Fermi energy as a function of the width $W$ of the conducting stripe which is proportional
to the number of occupied transversal modes: $n = k_{\rm F} W/\pi$ with Fermi wave number
$k_{\rm F}$. 
We see that both the total and the spin transmission increase linearly with $W$ 
yielding a constant polarization ratio $T_\mathrm{S}/T$ of the transmitted electrons. 
This ratio can be controlled by the strength of the SOI as shown in the inset 
of Fig.~\ref{HangMan}.\\
\noindent
While the linear increase of $T$ for large $W$ is expected and can be straightforwardly understood 
from the increasing number of open channels carried by the leads, the linear growth of $T_\mathrm{S}$
with wire width is remarkable.  In order to explain this linear rise of $T_\mathrm{S}$, 
we have to consider the number of SO-coupled channels participating in the spin polarization mechanism 
and to generalize the polarization mechanism described above to the case of sequences of
Landau-Zener-type transitions (see \cite{DA-Strehl} for an explicit treatment of the
case  $n\!=\!3$). Without invoking a further detailed
analysis of the SO-induced coupling mechanisms, a simple density-of-states argument shows that 
the number of relevant states in a critical energy window of order of the barrier height below the fixed
Fermi energy scales also linearly with $W$.
We finally note that, while the $W$-linear behavior of $T$ is also obtained classically,
the linear increase of $T_\mathrm{S}$ holds true for fully coherent, ballistic transport and 
presumably does not prevail for widths larger than the phase coherence length.
We have not explored its dependence on disorder.
\section{Spin ratchet effect}
After having explained the main mechanism responsible for spin polarization in dc-transport across
a single barrier, we are
now prepared to turn our attention to the more general case of a periodic arrangement of barriers
and ac-driving, i.e.\ the study of the spin ratchet 
 effect. 
We consider an adiabatic ac-driving assuming that the external bias 
$\mu _\mathrm{L}-\mu _\mathrm{R}$ is varied on a timescale much longer than the relevant 
timescales for electron transport through the system, which is basically the dwell time 
of the electrons in the ratchet scattering region. In experiments, the efficient operation 
of coherent charge ratchets in this adiabatic regime has already been confirmed~\cite{Linke99}. 

To be specific, here we consider an adiabatic square-wave driving with period $t_0$, 
where the chemical potential of the left/right reservoir,
\begin{equation}
\mu _{\mathrm{L/R}}(t)=
\begin{cases}
E_\mathrm{F} \pm U_0 / 2 \qquad\mathrm{for}\quad 0\leq t<t_0 / 2  \, ,    \\
E_\mathrm{F} \mp U_0 / 2 \qquad\mathrm{for}\quad t_0/ 2\leq t< t_0\,  , 
\end{cases}
\label{adrocking}
\end{equation}
%
is periodically switched [$\mu _{\mathrm{L}/\mathrm{R}}(t)=\mu _{\mathrm{L}/\mathrm{R}}(t+t_0)$]
between the two rocking situations with bias difference 
 $\pm U_0 = \mu _\mathrm{L}-\mu _\mathrm{R}$ with $U_0>0$.
For this adiabatic square wave driving, the ratchet is in a steady state 
in-between the switching events. Thus, we can use the respective expressions 
(\ref{Ic}) and (\ref{Is}) for the dc-charge  and spin current  
to calculate the averaged spin ratchet currents. For the driving of Eq.~(\ref{adrocking}) the
ratchet currents are given by the average between the two rocking conditions $+U_0$ and $-U_0$:
%
%
\begin{eqnarray}\label{Icnet}
\langle I (E_\mathrm{F} ,U_0)\rangle 
& = & \frac{1}{2}\left[I(E_\mathrm{F} ,+U_0)+I(E_\mathrm{F} ,-U_0)\right]\\
 & = &- \frac{e}{2h}\int_{0}^{\infty} \!\! \mathrm{d}E \;\Delta f(E;E_\mathrm{F} ,U_0) 
 \Delta T(E;U_0)  \, ,
 \nonumber
\end{eqnarray}
\begin{eqnarray}\label{Isnet}
\langle I^{\mathrm{S}}(E_\mathrm{F} ,U_0)\rangle & = & 
\frac{1}{2}\left[I^{\mathrm{S}}(E_\mathrm{F} ,+U_0)+ I^{\mathrm{S}}(E_\mathrm{F} ,-U_0)\right]\\
& = & \!\frac{1}{8\pi}\int_{0}^{\infty}\!\! \mathrm{d}E \;\Delta f(E;E_\mathrm{F} ,U_0)
\Delta T_\mathrm{S}(E;U_0) \, , \nonumber
\end{eqnarray}
%
where 
\begin{eqnarray}
\Delta f(E;E_\mathrm{F} ,U_0) & =&f(E;E_\mathrm{F} +U_0/2)  -  f(E;E_\mathrm{F} -U_0/2 ),\\ 
\Delta T(E;U_0)  & = &T(E;+U_0)-T(E;-U_0), \\
\Delta T_\mathrm{S}(E;U_0) &  =&T_{\mathrm{S}}(E;+U_0)-T_{\mathrm{S}}(E;-U_0). 
\label{ratspintrans}
\end{eqnarray}
%
In linear response [$\delta U_\mathrm{es}(x,y)=0$ in Eq.~(\ref{Hsim})] the quantities 
$\Delta T(E;U_0=0)=\Delta T_\mathrm{S}(E;U_0=0)$ vanish and thereby also the charge currents,
Eq.~(\ref{Icnet}) and spin ratchet currents, Eq.~(\ref{Isnet}). This implies that the
system has to be driven into the nonlinear regime, typical for ratchets.
Therefore, from now on we apply a finite driving bias $U_0$ to operate the spin ratchet. 
This bias determines the voltage drop $\delta U_\mathrm{es}(x,y)$ in Eq.~(\ref{Hsim}),
which describes the electrostatic potential due to the rearrangement of charges in the wire 
compared to the linear response case.

The spin ratchet mechanism, resulting from a finite $\Delta T_\mathrm{S}(E;U_0)$,
can be qualitatively understood from the Landau-Zener model used in Sec.~3
to explain the spin polarization mechanism of a single barrier. 
In \cite{Scheid2007}, the expression (\ref{LandauZener}) for the spin-flip probability
was extended to include a finite voltage drop:
\begin{equation}\label{LZRock}
P = 1-\exp \left\{ \frac{\eta}{(\partial/\partial x) 
(U_\mathrm{barr}(x)+\delta U_\mathrm{es}(x))} \right\}\,,
\end{equation}
where $\eta$  depends on the SOI strength and the confinement potential of the quantum wire,
 and
 the derivative is evaluated at the position $x$ of the avoided crossing marked by the dashed box
in Fig.~\ref{LZMod}.  We see that $P$ 
 depends on the amplitude of the driving voltage via the gradient 
$(\partial/\partial x) (U_\mathrm{barr}(x) +\delta U_\mathrm{es}(x))$ yielding different 
transition probabilities for forward and backward bias.
Since the transitions are induced when the degeneracy points (marked by the square in Fig.~\ref{LZMod}) cross the Fermi energy, the value of $\partial (\delta U_\mathrm{es})/\partial x$ in the vicinity of the barriers is important for the appearance of the spin ratchet effect. 
Therefore, this model predicts finite $\Delta T_\mathrm{S}$ for finite $U_0$, 
since  $P(+U_0)\neq P(-U_0)$. 
In summary, the spin ratchet effect predominantly results from the deformation of the 
potential barrier due to voltage drop $\delta U_\mathrm{es}$ which enters into the
SO-mediated spin-flip processes.

We first examine the ratchet spin current dependence on the number of transversal modes 
by approximating $\delta U_\mathrm{es}(x,y)$ by a linear function.
In Fig.~\ref{HangMan}b we see that, in analogy to the  dc-transport quantities $T$ and $T_\mathrm{S}$
(shown in Fig.~\ref{HangMan}a), the ratchet spin transmission $\Delta T_\mathrm{S}$ also exhibits
on average a linear increase for large $n$, though with a very small slope.
Its overall magnitude stays below one, and it exhibits larger values at small $n$ due to the strong
fluctuations in that regime. We can conclude that increasing the number of transverse modes
does not significantly enhance the ratchet spin current.

\subsection{Self-consistent calculation of the voltage drop}
In most models for ratchets, the effective voltage drop $\delta U_\mathrm{es}(x,y)$ along the wire
has been approximated by a linear function. In this section we will investigate the role of
the voltage drop by comparing spin ratchet signals resulting from
 $\delta U_\mathrm{es}(x,y)$ with its linear approximation.
To obtain $\delta U_\mathrm{es}(x,y)$ in the scattering region, we self-consistently solve 
the Schr\"odinger equation and the Poisson equation. 
To this end we adopt the approach introduced for ratchets in \cite{Scheid2009a}, which we 
describe in the following.
Within this approach we absorb all electrostatic potentials (e.g.~the potential due to donor atoms),
which do not change upon variation of system parameters, as do the Fermi energy or the bias voltage, in
the confinement potential of the wire. Then we only have to consider the rearrangement of the electrons,
$\delta n = n-n_0$, due to a finite bias in order to determine the voltage drop in the system.\\
For the calculation of the electron density we determine the lesser Green's function
$\mathcal{G}^<$ of the system via the Keldysh equation~\cite{B:Haug}
\begin{equation}\label{G_Less_Scatt}
\mathcal{G}^<=\mathcal{G}^\mathrm{r}\Sigma ^< \left( \mathcal{G}^\mathrm{r}\right) ^\dagger\,,
\end{equation}
where $\mathcal{G}^\mathrm{r}$ is the retarded Green's function of the system, which we 
evaluate via a recursive Green's function method~\cite{Wimmer2009}. Furthermore, the lesser self-energy $\Sigma ^<$ can 
be expressed in terms of the retarded self-energies $\Sigma ^\mathrm{r}_{\mathrm{L}_i}$ of the individual leads $i$,
\begin{equation}\label{Sigm_Less}
\Sigma ^< =-\sum _i f(E,\mu _i)\left[ \Sigma ^\mathrm{r}_{\mathrm{L}_i} - \left( \Sigma ^\mathrm{r}_{\mathrm{L}_i}\right)^\dagger \right]\,.
\end{equation}
Finally, the electron density of the system is given by
\begin{equation}\label{DensitiesC}
n(\vec{r})=-\frac{\rm{i}}{2\pi}\int_{-\infty}^\infty \mathrm{d}E \,\mathrm{Tr}[\mathcal{G}^<(\vec{r},\vec{r};E)]\, .
\end{equation}
We then obtain the change in the electrostatic potential, $\delta U_{\rm es} = U_{\rm es}-U^0_{\rm
es}$, by solving the corresponding Poisson equation~\cite{Xue2002}
\begin{equation}
  \label{eq:poisson}
  \vec{\nabla}^2 \delta U_{\rm es}(\vec{r}) = 
  -\frac{e^2}{\varepsilon_r \varepsilon_0} \Big[n(\vec{r}) - n_0(\vec{r})\Big] \,,
\end{equation}
where $\varepsilon_0$ is the vacuum permittivity and $\varepsilon_r$ is the material specific relative static permittivity.
For the evaluation of $\delta U_{\rm es}(\vec{r})$ it is useful to distinguish
the contributions from the leads, $\delta U_{\rm lead}(\vec{r})$, and 
from the rearrangement of the electrons in the scattering region, $\delta U_{\rm sr}(\vec{r})$~\cite{Xue2002,Nitzan2002}:
\begin{equation}
\delta U_{\rm es}(\vec{r})=\delta U_{\rm lead}(\vec{r})+\delta U_{\rm sr}(\vec{r})
\end{equation}
Here, $\delta U_{\rm lead}(\vec{r})$ solves the Laplace equation $\vec{\nabla}^2 \delta U_{\rm lead}(\vec{r})
= 0$ with the boundary conditions $\delta U_\mathrm{lead}(x=\pm L/2) = \mp U_0/2$
and is therefore given by a linear function between both contacts~\cite{Nitzan2002}.
On the other hand, $\delta U_{\rm sr}(\vec{r})$ solves 
the Poisson Eq.~(\ref{eq:poisson}) with boundary conditions $\delta U_\mathrm{sr}(x=\pm L/2) = 0$. To obtain 
$\delta U_{\rm sr}(\vec{r})$ from this equation we 
assume that the electron density inside the leads is much higher than in 
the scattering region. In the simulations we realize this by introducing an additional electrostatic offset potential 
in the scattering region, see e.g. Fig.~\ref{SORat}d). As a result the electrostatic potential profile close to the 
leads is flat, i.e.~$n(\vec{r}) \approx n_0(\vec{r})$, which enables us to calculate $\delta U_{\rm sr}$ from 
the Poisson Eq.~(\ref{eq:poisson}) with vanishing $\delta U_{\rm sr}$ for $|\vec r|\rightarrow\infty$, 
yielding
\begin{equation}
  \label{eq:V_el}
  \delta U_{\rm sr}(\vec{r}) = \frac{e^2}{4\pi\varepsilon_r \varepsilon_0} \int {\rm d}^2
  r' \frac{n(\vec{r'}) - n_0(\vec{r'})}{|\vec{r}-\vec{r'}|}\,.
\end{equation}
Now we can compute the 
electrostatic potential $\delta U_{\rm es}(x,y)$ at finite bias voltages. To this end,
we start with an initial guess for $\delta U_{\rm es}(x,y)$ and calculate 
the electron density $n(x,y)$ for this case via Eq.~(\ref{DensitiesC}). 
By solving Eq.~(\ref{eq:V_el}) we obtain a new potential $\delta U_{\rm es}(x,y)$, which in turn 
can be used to calculate the corresponding density $n(x,y)$. This procedure is repeated until 
convergence is reached.
In practice, we do not directly iterate between Eqs.~(\ref{DensitiesC}) and (\ref{eq:V_el}), but we use 
the so-called Newton-Raphson method, which has been successfully applied to similar non-equilibrium problems~\cite{Trellakis1997,Lake1997} and significantly improves the convergence of the self-consistent calculations.\\
Before turning to the spin current calculation
we first present a symmetry analysis of the spin-resolved 
transmission probabilities~\cite{Scheid2007a,Zhai2005}. This is helpful to simplify the expressions for the ratchet currents.  To be specific, the Hamiltonian (\ref{Hsim}) is invariant 
under the operation of
\begin{equation}\label{ACSymRel}
\hat{\mathcal{P}}=-\mathrm{i}\hat{\mathcal{C}}\hat{R}_\mathrm{U}\hat{R}_x\sigma _z \, , 
\end{equation}
since the symmetry relations
\begin{equation}\label{ratsym}
\delta U_\mathrm{es}(x)=- \delta U_\mathrm{es}(-x),\; U_\mathrm{barr}(x)=U_\mathrm{barr}(-x),\;\alpha (x)=\alpha (-x)
\end{equation}
are fulfilled. In Eq.~(\ref{ratsym}), 
$\hat{R}_x$ inverts the $x$-coordinate, $\hat{R}_\mathrm{U}$ switches the sign of
the bias voltage ($\pm U_0\leftrightarrow \mp U_0$) and $\hat{\mathcal{C}}$ is the operator of complex
conjugation. The first equality, $\delta U_\mathrm{es}(x)=- \delta U_\mathrm{es}(-x)$, is due to the
spatial symmetries of the system and the use of the same computational scheme for both forward and 
backward bias. 
In \cite{Scheid2007a} we showed that due to this invariance the relation 
\begin{equation}\label{TrelRat}
T_{\sigma ,\sigma '}(E,\pm U_0)=T_{\sigma ', \sigma}(E,\mp U_0)
\end{equation}
is fulfilled. As a consequence, the charge ratchet current $\langle I\rangle$, Eq.~(\ref{Icnet}), for this 
system vanishes. On the other hand, the expression for the spin ratchet current, Eq.~(\ref{Isnet}), can be simplified through the above symmetry relation:
\begin{equation}
\langle I^{\mathrm{S}}(U_0)\rangle =\frac{1}{4\pi}\int_{0}^{\infty} \mathrm{d}E \;\Delta f(E;U_0)\Big[T_{+,-}(E,+U_0)-T_{-,+}(E,+U_0)\Big] \,.
\end{equation}
Hence, it is sufficient to calculate the spin-flip transmission probabilities $T_{+,-}$ and $T_{-,+}$ for a single rocking condition.\\
\begin{figure}[!tb]
\begin{center}
 	\includegraphics[width=0.8\linewidth]{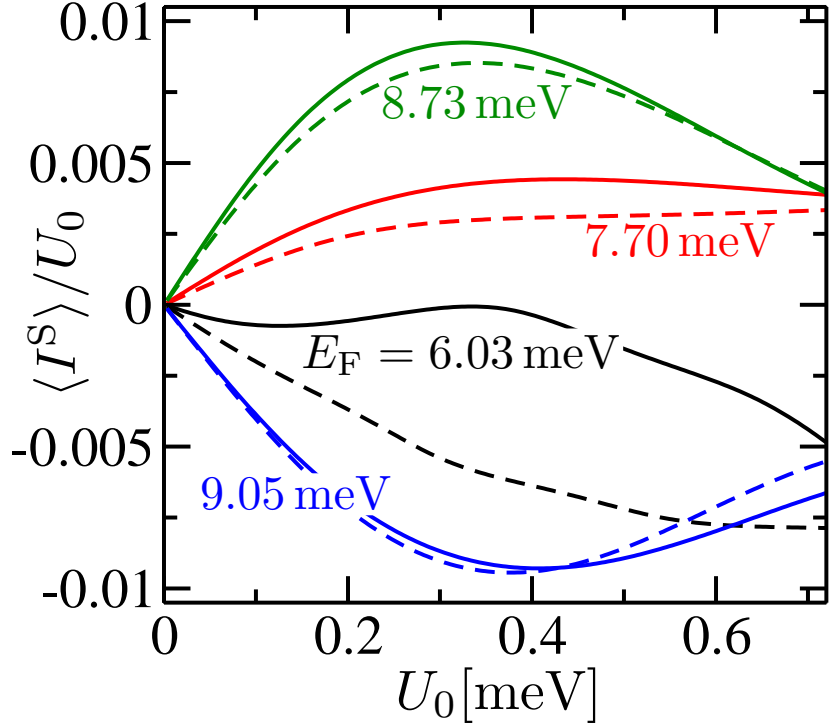}
  \caption{Ratchet spin conductance $\langle I^\mathrm{S}\rangle/U_0$ as a function of the driving
voltage $U_0$ for four different values of $E_\mathrm{F}$. We
compare the results for a linear voltage drop (dashed lines) with those for the self-consistently determined 
$\delta U_\mathrm{es}(x,y)$ (solid lines). Parameters: $L_\mathrm{B}=100$ nm, $N_\mathrm{B}=5$, $\alpha _0 =4.76\cdot 10^{-11} \,\mathrm{eV}\,\mathrm{m}$, $W=250$ nm, $W_0=150$ nm, $U_\mathrm{B}=3.17$ meV, $U_\mathrm{os}=2.38$ meV.
  \label{SC_DC}}
\end{center}
\end{figure}\noindent
In Fig.~\ref{SC_DC} we present the calculated spin ratchet conductance
 $\langle I^\mathrm{S}\rangle /U_0$ as a function of the amplitude 
of the driving voltage $U_0$, see Eq.~(\ref{adrocking}). As the 2DEG material we choose InAs with the parameters 
$m^*=0.024 \,m_0$, $g^*=15$ and 
$\varepsilon _\mathrm{r}=15.15$. Furthermore, the lattice spacing is set to $a=10\,\mathrm{nm}$.
We find a finite spin current with a direction depending on the Fermi energy.
We compare the results of a linear voltage drop between $x=-L/2$ and $x=L/2$ (dashed 
curves in Fig.~\ref{SC_DC}) with those obtained from the self-consistent calculation of 
$\delta U_\mathrm{es}(x,y)$ (solid curves) for several representative Fermi energies. 
The self-consistent calculations and the linear voltage drop model yield similar results for 
$\langle I^\mathrm{S}\rangle$ except for the case $E_\mathrm{F} =6.03$ meV,
where transport is dominated by resonant tunneling and hence depends sensitively on the 
details of the potential.
\begin{figure}[!tb]
\begin{center}
 	\includegraphics[width=0.5\linewidth]{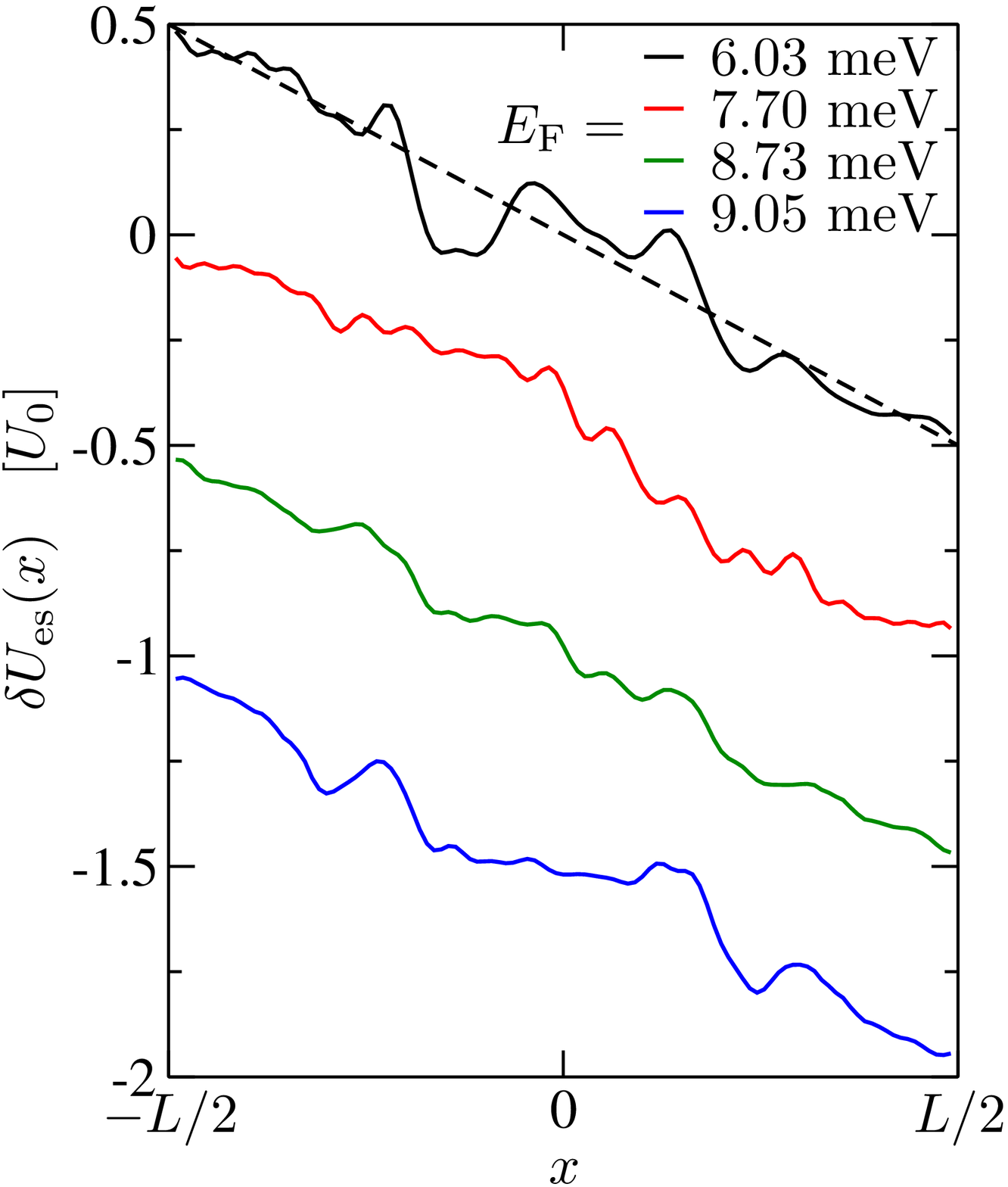}
  \caption{Form of the self-consistently determined voltage drop $\delta U_\mathrm{es}(x,y)$ averaged over the transversal $y$-direction for four different values of $E_\mathrm{F}$. The results shown were obtained for the system investigated in Fig.~\ref{SC_DC} at bias $U_0=0.24$ meV. For clarity, the curves are vertically offset by a value of $-0.5U_0$ each. For comparison, the dashed line shows the linear voltage drop considered in Fig.~\ref{SC_DC}.
  \label{SC_VD}}
\end{center}
\end{figure}\noindent
In Fig.~\ref{SC_VD} we depict the spatial distribution of the self-consistently calculated voltage 
drop $\delta U_\mathrm{es}$ along the ratchet wire. We see that the linear ramp is indeed a good 
approximation for the three higher Fermi energy values considered. 
This also explains the good qualitative agreement in Fig.~\ref{SC_DC} between the spin conductance
for the self-consistently determined $\delta U_\mathrm{es}$ and the linear voltage drop model, respectively.
For the case of $E_\mathrm{F} =6.03$ meV the resulting voltage drop shows the most pronounced non-monotonic 
behavior in Fig.~\ref{SC_VD}. 
This causes a misalignment of the energy levels in the potential valleys and therefore a 
reduction of the miniband-mediated resonant transport. As a consequence the spin ratchet current is 
overestimated in the linear voltage drop model for this value of  $E_\mathrm{F}$.
\subsection{Influence of disorder}

\begin{figure}[!tb]
\begin{center}
 	\includegraphics[width=0.8\linewidth]{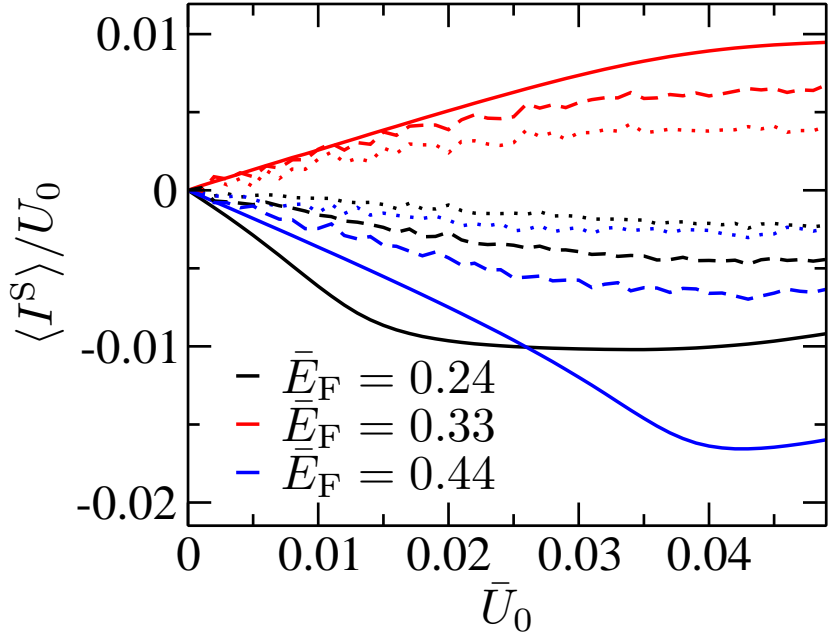}
\caption{Spin ratchet conductance $\langle I^\mathrm{S}\rangle/U_0$ as a function of the driving
amplitude $\bar{U}_0$ for three different scaled Fermi energies, $\bar{E}_\mathrm{F}=0.24$ (black lines), $\bar{E}_\mathrm{F}=0.33$ (red lines), $\bar{E}_\mathrm{F}=0.44$ (blue lines). Results are shown for a clean system (solid lines) and for systems with Anderson disorder of strength $\bar{U}_\mathrm{dis}=0.5$ (dashed lines) and $\bar{U}_\mathrm{dis}=\sqrt{2}/4$ (dotted lines).
Parameters: $N_\mathrm{B}=5$, $L_\mathrm{B}=10a$, $\bar\alpha _0 =0.15$, $W=W_0=15a$, $\bar U_\mathrm{B}=0.22$.
  \label{SOrat:10b}}
\end{center}
\end{figure}

So far we have studied the case of disorder free, clean conductors. 
However, in realistic experimental samples dopands, crystal defects and impurities 
give rise to momentum scattering. This in turn can cause spin relaxation~\cite{Dyakonov1972}, 
which might limit the performance of the spin ratchet. 
In order to investigate the role of impurity scattering on the ratchet effect
we again consider the device shown in Fig.~\ref{SORat} but
 with $W=W_0=15a$ and without the additional potential offset shown in Fig.~\ref{SORat}d. 
For  the sake of computational feasibility in the following we assume a linear voltage drop 
in the central region, since we have seen that such a model represents a fair approximation 
for the actual voltage drop for a wide parameter range,
 and for the sake of computational feasibility. For a fixed bias $U_0$, it is given by
\begin{equation}\label{HeurSOVDrop}
\delta U^\mathrm{lin}_\mathrm{es}(x)=\begin{cases}
U_0/2 &\mathrm{for}\;x<-N_\mathrm{B}L_\mathrm{B}/2 \\
U_0 x/(N_\mathrm{B}L_\mathrm{B})&\mathrm{for}\;|x|<N_\mathrm{B}L_\mathrm{B}/2\\
-U_0/2 &\mathrm{for}\;x>N_\mathrm{B}L_\mathrm{B}/2 
\end{cases}\,.
\end{equation}
In Fig.~\ref{SOrat:10b} we present the spin ratchet conductance $\langle I^\mathrm{S}\rangle/U_0$ as 
a function of the scaled driving amplitude $\bar{U}_0$ for three representative Fermi energies. 
In view of Eq.~(\ref{LZRock}), we expect the spin ratchet mechanism to be enhanced for higher 
driving amplitudes $\bar{U}_0$, since the difference between $P(+U_0)$ and 
$P(-U_0)$ depends on the slope of the voltage drop. 
Indeed, in Fig.~\ref{SOrat:10b} we find a linear increase of the spin ratchet conductance for 
small $\bar{U}_0$, which can be understood by expanding  Eq.~(\ref{LZRock}) in this limit. \\
\noindent
We model impurity scattering by introducing Anderson disorder in the tight-binding version of the 
Hamiltonian~(\ref{Hsim}). To this end we add a random potential from the box distribution
$[-U_\mathrm{dis}/2;U_\mathrm{dis}/2]$ to the on-site energy of each site in the central region 
of the wire.
In Fig.~\ref{SOrat:10b} we compare the case of the clean quantum wire (solid lines) 
with results for the spin ratchet conductance for a disordered quantum wire (dashed and dotted lines). 
Although $\langle I^\mathrm{S}\rangle /U_0$ is reduced compared to the disorder-free system, 
the surviving spin ratchet conductance still possesses a reasonable magnitude. \\
In order to get an idea of the corresponding mean free path in those calculations, 
we now exemplary set the lattice spacing to $a=10$nm and choose InAs with $m^*=0.024\,m_0$ as the 
2DEG material. Then the width of the ratchet device is $W=W_0=150\,\mathrm{nm}$, 
the length of a single barrier is $L_\mathrm{B}=100\,\mathrm{nm}$ and hence the overall 
length of the ratchet set-up is of order 1 micron. The Rashba SO strength is 
$\alpha _0 \approx 4.76\cdot 10^{-11} \,\mathrm{eV}\,\mathrm{m}$, a value well in reach of present day 
experiments~\cite{Grundler2000}. 
The mean free path in this system can then be approximated as 
$l=48a\sqrt{\bar{E}_\mathrm{F}}/\bar{U}_\mathrm{dis}^2$~\cite{Ando1991}. 
Then the disorder strengths chosen in Fig.~\ref{SOrat:10b}, 
$\bar U_\mathrm{dis}=1/2$ and $\sqrt{2}/4$,
correspond to an elastic mean free path of the order of one and two microns, respectively,
for all Fermi energies considered. 
We hence can conclude from Fig.~\ref{SOrat:10b}
that the spin ratchet effect prevails if $l$ is of the order of or larger than the ratchet 
system size, i.e.\ even in the disordered but non-diffusive limit.
Since  mean free paths $l \gg 1\mu$m are possible in clean InAs quantum wells~\cite{Koester1996}, 
in realistic experimental situations spin ratchet 
output signals of reasonable magnitude should be observable.

\section{Conclusions}

In this paper, one the one hand, we have considered the spin-dependent transmission
through single and multiple barriers in stripes build from two-dimensional
electron gases with Rashba SO interaction. We have shown
that for coherent transport across a smooth barrier with SO-mixed transverse
channels, the spin transmission, i.e.\ difference between transmitted particles
of opposite spin direction, increases linearly with channel number $n$.
This implies an in-plane polarization of
transmitted charge carriers, which is independent of $n$ or
the wire width, respectively, and thereby distinctly larger than
usual mesoscopic spin transmissions, such as e.g.\ from
conductance fluctuations \cite{Sharma03,Krich} or in the context of the 
mesoscopic spin Hall effect \cite{Barderson} which are typically of order 1.

On the other hand, we have addressed the SO-mediated spin ratchet mechanism
and thereby extended previous work, which gave a proof of principle for
SO ratchets \cite{Scheid2007}, to realistic mesoscopic systems by accounting
for certain nonequilibrium and disorder effects. In particular, we
presented a self-consistent treatment of the ratchet transport in
the strongly nonlinear, though coherent regime, showing that linear
voltage-drop models often represent good approximations to the real
charge rearrangement but may break down in parameter regimes dominated
by resonant charge and spin transfer. We further investigated the effects
of elastic impurity scattering due to static disorder. We found that,
as expected, the ratchet spin current decreases with increasing scattering
rate, but a finite fraction of the clean spin current prevails (with the
same sign) if the elastic mean free path is of the order of or larger than 
the system size.

We close with a few further remarks:
For systems with bulk inversion asymmetry, such as GaAs,
Dresselhaus SO interaction  \cite{Dresselhaus} has additionally to be considered.
Calculations~\cite{DA-Pfund} for a ballistic SO ratchet with combined
Rashba- and Dresselaus SO coupling showed that the overall
picture is not altered and furthermore demonstrate that the ratchet
spin current direction can be changed upon tuning the relative strength
of the two SO coupling mechanisms. In particular one finds spin current
reversals for equal Rashba- and Dresselaus SO interaction where the
SO effects cancel.
This feature allows for inverting the polarization direction of the
output current by simple electrical means, namely by tuning the
Rashba SO strength, e.g.\ through a gate voltage.


By now we have reached a fairly complete picture
of the  SO-based spin ratchet mechanism in both
limits, the quantum coherent and the strongly dissipative regime.
A further extension of previous work and a future challenge would 
consist in bridging these two separate limits treated up to now.

In the coherent regime we so far considered the case of
adiabatic driving that is relevant, e.g., for ratchet experiments
employing nanostructures with an external ac bias voltage.
If the system is driven through external radiation, the timescales
for driving and for electron or spin dynamics can be comparable,
which may give rise to further interesting spin ratchet phenomena. This ac regime
requires an approach beyond the adiabatic limit, e.g.\ a Floquet treatment.

The concept to extend the particle ratchet mechanism to other
quantities such as the spin degree of freedom may be further generalized.
Along this line one may think of extending this concept to other 
spin 1/2-type quantities, for instance the pseudo- or valley-spin degree of
freedoms of charge carriers in graphene. More generally one may
think of devising ratchet mechanisms to spatially separate objects
according to their further internal degrees of freedom, for instance
atoms with internal two- or multi-level dynamics.

\section{Acknowledgements} 

We thank P.~H\" anggi for numerous
interesting discussions on ratchets, transport physics and beyond,
and continuous support throughout many years.
We thank I.~Adagideli, M.~Grifoni, A.~Lassl, A.~Pfund, S.~Smirnov, M.~Strehl
and M.~Wimmer for useful conversations. We acknowledge funding from DFG through
SFB 689. MS acknowledges further support from the
 {\em Studienstiftung des Deutschen Volkes}. The work of DB  
was partially supported 
by the Excellence Initiative of the German Federal and State Governments.

\end{document}